\documentclass{iau} 
\usepackage{graphicx}
\usepackage{epstopdf}
\usepackage{natbib}
\usepackage[percent]{overpic}
\usepackage{color}

\title[Ca II K spectroheliograms for studies of long-term changes in solar irradiance]{Ca II K spectroheliograms for studies of long-term changes in solar irradiance}

\author[T. Chatzistergos, I. Ermolli, N. A. Krivova, \& S. K. Solanki]   
{Theodosios Chatzistergos$^1$, Ilaria Ermolli$^1$, Natalie A. Krivova$^2$, \& Sami K. Solanki$^{2,3}$}

\affiliation{
$^1$INAF Osservatorio Astronomico di Roma, Monte Porzio Catone, Italy\\[\affilskip] 
$^2$Max-Planck-Institut f\"{u}r Sonnensystemforschung, G\"{o}ttingen, Germany \\[\affilskip]
$^3$School of Space Research, Kyung Hee University, Yongin,  Republic of Korea}

\pubyear{2018}
\volume{340}  
\jname{Long--Term Datasets for the Understanding of \\ Solar and Stellar Magnetic Cycle}
\editors{}
\begin{document}

\maketitle

\begin{abstract}
We address the importance of historical full disc Ca II K spectroheliograms for solar activity and irradiance reconstruction studies. We review our work on processing such data to enable them to be used in irradiance reconstructions. We also present our preliminary estimates of the plage areas from five of the longest available historical Ca II K archives.
\end{abstract}

\firstsection 
              
\section{Introduction}
Solar observations in the Ca II K line (393.34 nm) started as early as 1892 \citep{hale_solar_1893}, while more systematic surveys of the Sun in this line have been pursued since 1907 \citep[see also][]{ermolli2018_proc}. Such data continue to be collected till the present time at numerous observatories around the globe  providing a good coverage of the whole 20th century \citep[see][for a list of the various available Ca II K archives]{chatzistergos_analysis_2017}.
These data describe the evolution of the plage regions, and thus carry invaluable information on the past solar magnetism \citep[see e.g.][]{schrijver_relations_1989,loukitcheva_relationship_2009,pevtsov_reconstructing_2016,kahil_brightness_2017}. Consequently, Ca II K observations can be used as input to irradiance models \citep[e.g.][]{ermolli_modeling_2003,krivova_reconstruction_2003,krivova_reconstruction_2010,fontenla_signature_2004,fontenla_high-resolution_2011,yeo_reconstruction_2014} to potentially allow more accurate reconstructions over the last century, as well as to impose constraints on reconstructions going further back in time. 

Historical Ca II K observations have so far not been employed in irradiance reconstructions, mostly because the images lack proper photometric calibration and suffer from multiple artefacts.
\cite{chatzistergos_analysis_2018} have presented a new method to accurately calibrate the images and compensate for the intensity limb darkening and several image artefacts. The method was exhaustingly tested, showing its efficiency. Also, the limb darkening is removed more accurately than in previous studies presented in the literature. Here we apply this method on data from five of the longest currently available historical Ca II K archives and present preliminary results for the plage area evolution.

\section{Data and methods}
\begin{figure*}
	\centering
	\includegraphics[width=1\linewidth]{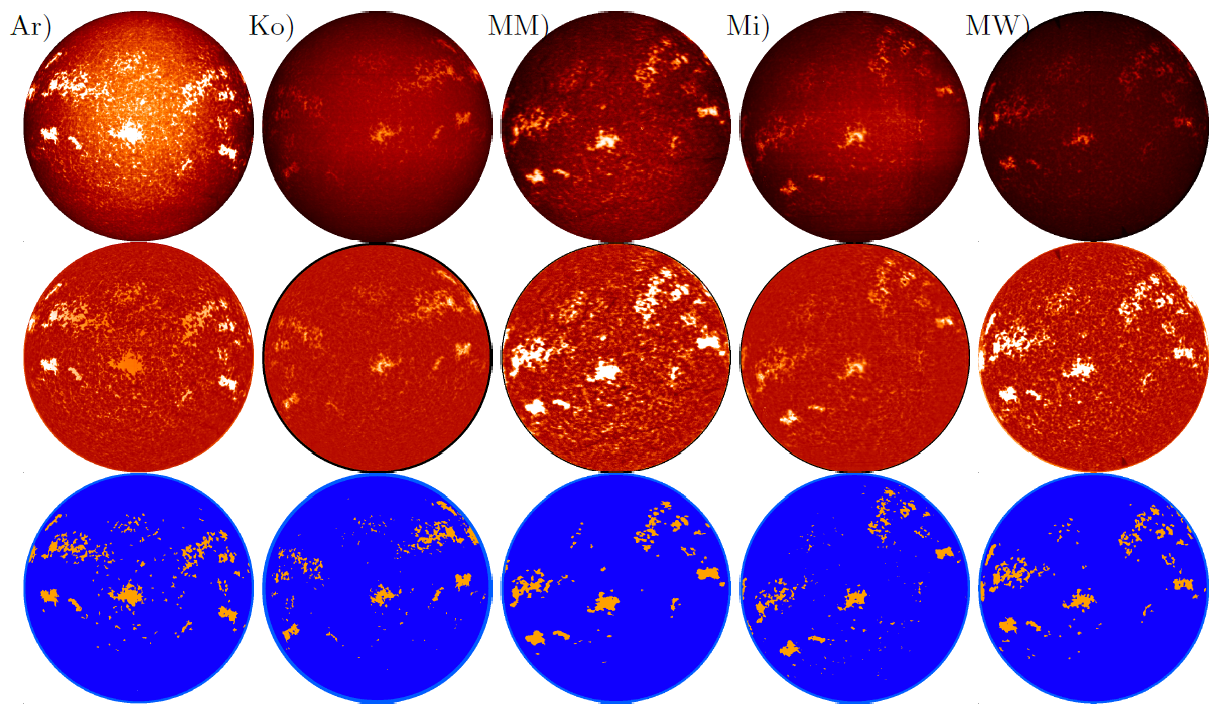}
	\caption{Examples of our processing of the Ca II K observations taken on 05/05/1959. Shown are the raw density image (top row), photometrically calibrated and limb-darkening-compensated image (middle row), and segmentation mask (bottom row). Each column shows the results for a different archive, from left to right: Arcetri, Kodaikanal, McMath-Hulbert, Mitaka, and Mt Wilson. All density images are normalised to the range of values within the disc, while the limb-darkening-compensated images are shown in the range [-0.5,0.5]. The plage regions in the mask are shown in orange, while the quiet Sun regions (including all dark regions such as sunspots or filaments) are shown in blue. The images are not compensated for ephemeris.}
	\label{fig:processedimagessamedayoriginal}
\end{figure*}

\begin{figure}
	\centering
	\includegraphics[width=0.75\linewidth]{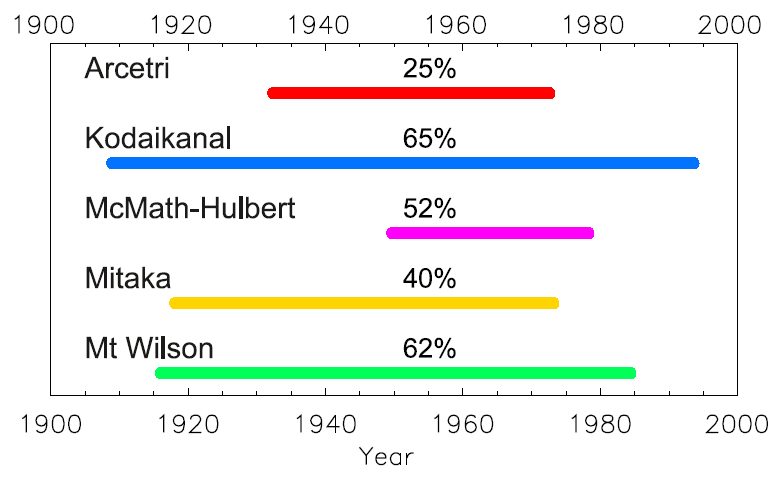}
	\caption{Temporal coverage of the five Ca II K archives used in this study. The percentage of days with at least one observation within the given intervals is also shown.}
	\label{fig:periods2}
\end{figure}

We analysed full-disc Ca II K spectroheliograms from the digitised archives of Arcetri \cite[Ar,][]{ermolli_comparison_2009}, Kodaikanal \cite[Ko,][]{makarov_22-years_2004}, McMath-Hulbert \cite[MM,][]{fredga_comparison_1971}, Mitaka \cite[Mi,][]{hanaoka_long-term_2013}, and Mt Wilson \cite[MW,][]{lefebvre_solar_2005} observatories.  
The top row of Fig. \ref{fig:processedimagessamedayoriginal} shows digitised Ca II K images from all archives taken on the same day. The bandwidth of the observations is between 0.01 and 0.05 nm. 
Figure \ref{fig:periods2} shows the temporal coverage of the five archives. Ko is the longest dataset with the best coverage, however MW currently has more observations available in digital form. 

The photometric calibration and compensation of the limb darkening of the images was done following \cite{chatzistergos_exploiting_2016,chatzistergos_analysis_2018} and \cite{chatzistergos_analysis_2017}. Examples of the processed images can be seen in the middle row of Fig. \ref{fig:processedimagessamedayoriginal}.
The contrast of the MM and MW images is greater than that in Ar, Ko, and Mi images, which is consistent with the narrower bandwidth used at the former observatories compared to the latter ones. We also see saturated regions in the Ar observation \cite[see][]{chatzistergos_analysis_2017}.
 
The images were then segmented with a constant multiplicative factor to the standard deviation of the contrast of the quiet Sun regions over the disc. The multiplicative factor used to identify plage regions was chosen to be $m=3.25$ and we used the same value for images from all the different archives. Example segmentation masks are shown in the bottom row of Fig. \ref{fig:processedimagessamedayoriginal}.

\section{Plage areas}
\begin{figure}
	\centering
	\includegraphics[width=1\linewidth]{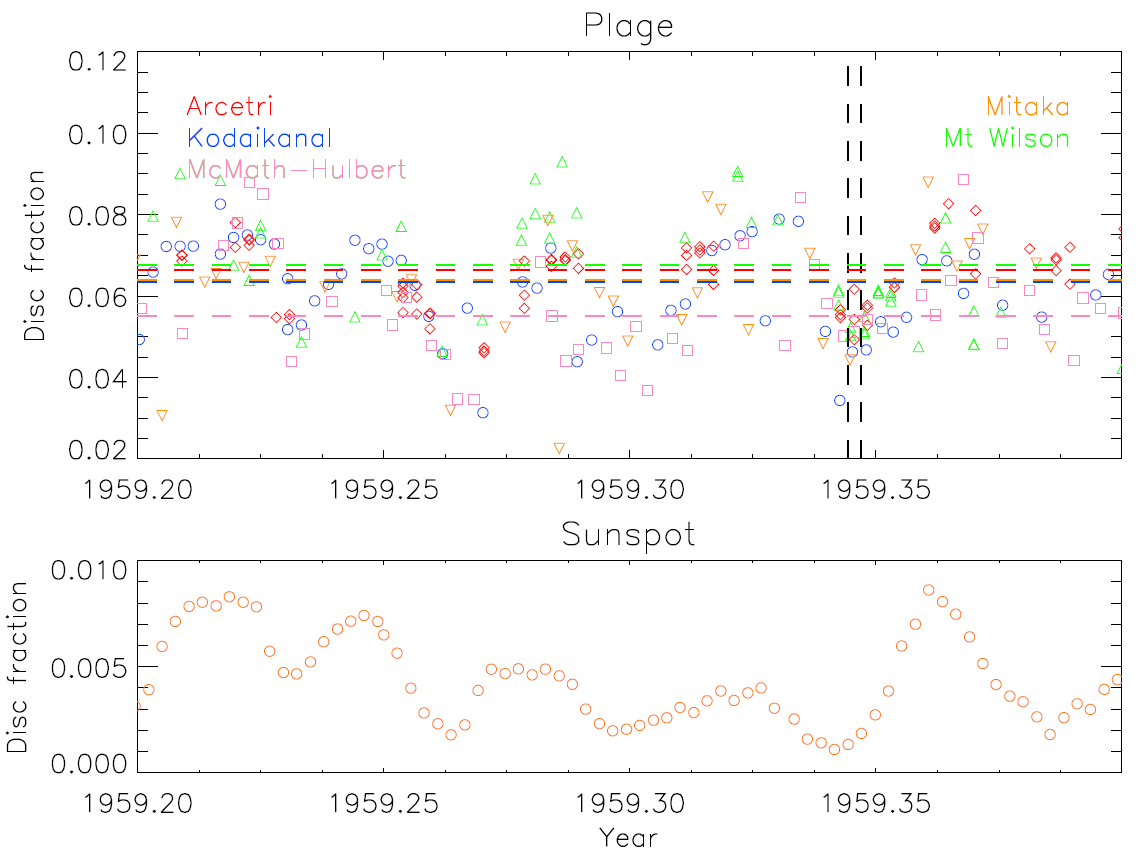}
	\caption{Top: Plage areas derived from the observations at the Arcetri (red rhombuses), Kodaikanal (blue circles), McMath-Hulbert (pink squares), Mitaka (orange downward triangles), and Mt Wilson (green upward triangles) observatories over a 74-day period around the activity maximum of solar cycle 19. The dashed coloured horizontal lines represent the median value of the plage areas for each archive. The dashed black vertical lines enclose the day of the observations shown in Fig. \ref{fig:processedimagessamedayoriginal}. Bottom: the sunspot areas from \cite{balmaceda_homogeneous_2009} over the same period.}
	\label{fig:1discfractionplage}
\end{figure}

Figure \ref{fig:1discfractionplage} shows the derived plage areas from all five historical archives used here over a 74-day period at the beginning of 1959, which is near the maximum of solar cycle 19. The sunspot areas from \cite{balmaceda_homogeneous_2009}\footnote{Available at http://www2.mps.mpg.de/projects/sun-climate/data.html} are shown in the bottom panel. 

Over the selected period the plage areas derived from all archives show a similar temporal variability pattern. The results from Ar and Ko match particularly well. The MM areas are generally somewhat lower than the others, while following the overall profile of the variability. Note that here we used the same constant parameter for the segmentation of all archives, which is rather simplistic.
The differences in the temporal coverage by the different archives is also seen.
For instance, Ko samples more days than the other archives over the period considered. However, Ko has single observations per day while Ar and MW have multiple observations for most of these days.

\section{Summary}
Historical Ca II K data bear significant potential for studies of long-term changes in solar activity and irradiance. \cite{chatzistergos_analysis_2018} have recently developed a method to perform the photometric calibration, remove the limb darkening, and account for various artefacts. Here we present results of the analysis of a roughly 2-month long sample of Ca II K images from five historical archives. We show that results from the various archives are consistent and show a similar temporal variation. A more detailed study of the complete datasets will be presented in a forthcoming paper.

\begin{acknowledgements}
	We thank the Arcetri, Kodaikanal, McMath-Hulbert, Mitaka, Mount Wilson, and the Rome Solar Groups.
	T.C. acknowledges postgraduate fellowship of the International Max Planck Research School on Physical Processes in the Solar System and Beyond.
	This work was supported by grants PRIN-INAF-2014 and PRIN/MIUR 2012P2HRCR "Il Sole attivo", COST Action ES1005 "TOSCA", FP7 SOLID, and by the BK21 plus program through the National Research Foundation (NRF) funded by the Ministry of Education of Korea.
\end{acknowledgements}

\end{document}